\documentclass{article}
\pdfoutput=1

\usepackage{authblk}

\usepackage{times}  
\usepackage{epsfig}
\usepackage[TABBOTCAP]{subfigure}
\usepackage{tabularx}
\usepackage{graphicx} 
\usepackage{color}
\usepackage{xspace}
\usepackage{thumbpdf}
\usepackage{listings}
\usepackage{verbatim}
\usepackage{booktabs}
\usepackage{colortbl}

\usepackage{amsmath,amssymb}
\usepackage{color}
\usepackage{tikz}
\usetikzlibrary{plotmarks,shapes,snakes}
\usepackage{pgfplots}
\usepackage{hyperref}
\usepackage{multicol}
\usepackage{listings}

\definecolor{gray}{rgb}{0.4,0.4,0.4}
\definecolor{darkblue}{rgb}{0.0,0.0,0.6}
\definecolor{cyan}{rgb}{0.0,0.6,0.6}

\lstdefinelanguage{XML}
{
  basicstyle=\ttfamily\small,
  morestring=[b]",
  morestring=[s]{>}{<},
  morecomment=[s]{<?}{?>},
  stringstyle=\color{black},
  identifierstyle=\color{darkblue},
  keywordstyle=\color{cyan},
  morekeywords={xmlns,version,type}
}

\usepackage{paralist}
\usepackage{cite}
\usepackage{color}
\usepackage{subfigure}
\usepackage{times}
\usepackage{epsfig}
\usepackage{graphics}
\usepackage{float}
\usepackage{color}
\usepackage{url}
\usepackage{algorithm}
\usepackage{algpseudocode}
\usepackage{multirow}
\usepackage{icomma}
\usepackage{fullpage}

\def\legendsize{\scriptsize}

\definecolor{lightblue}{RGB}{200,200,255}
\definecolor{lightgreen}{RGB}{200,255,200}
\definecolor{darkgreen}{RGB}{100,155,100}

\newcommand{\barch}{BASEL}

\hypersetup{pdfstartview=FitH,pdfpagelayout=SinglePage}

\begin{document}


\title{\barch{} (Buffering Architecture SpEcification Language)}
\author[1]{Kirill Kogan}
\author[2]{Danushka Menikkumbura}
\author[3]{Gustavo Petri}
\author[4]{Youngtae Noh}
\author[5]{Sergey Nikolenko}
\author[2,6]{Patrick Eugster}
\affil[1]{IMDEA Networks Institute}
\affil[2]{Purdue University}
\affil[3]{Universit\'e Paris Diderot - Paris 7}
\affil[4]{Inha University}
\affil[5]{Steklov Mathematical Institute}
\affil[6]{TU Darmstadt}


%
%
%
%
%
%
\maketitle


\subsection*{Abstract}
Buffering architectures and policies for their efficient management constitute one of the core ingredients of a network architecture. In this work we introduce a new specification language, \barch{}, that allows to express virtual buffering architectures and management policies representing a variety of economic models. \barch{} does not require the user to implement policies in a high-level language; rather, the entire buffering architecture and its policy are reduced to several comparators and simple functions. We show examples of buffering architectures in \barch{} and demonstrate empirically the impact of various settings on performance.

\section{Introduction}


Design and management of a \emph{buffering architecture} are key elements in meeting network design challenges since they directly impact the performance and cost of each network element.
Application-induced traffic bursts can create an imbalance between incoming and outgoing packet rates to a given port, so packets must be queued in the network element. The available queue size on a port determines the port's ability to hold packets until the egress port can emit them. Packets are dropped when buffer resources are congested,  resulting in poor performance. The allocation and availability of buffer resources to ports is determined not only by the buffer's size but also by the buffering architecture of the network element. Overprovisioning in terms of buffer capacity at each network node to absorb bursty behavior is not viable, as networks cannot have unlimited resources; on the contrary, data centers can only scale out as fast as the effective per-port cost and power consumption. These factors are, in turn, defined by the chosen buffering architecture.

Objectives beyond \emph{fairness}, and the incorporation of additional traffic properties, lead to new challenges in the implementation and performance for traditional switching architectures~\cite{nofair,Gettys13,JainKMOPSVWZZZHSV13,HongKMZGNW13}. This calls for novel abstractions that enable the definition of buffering architectures and management policies that can be deployed on real network elements. Designing such abstractions however is non-trivial, as they must satisfy a number of possibly conflicting requirements: 
\begin{inparaenum}
\item[(1) {\sc Expressivity}:] expressible policies should cover a large majority of buffering architectures representing common economical models of existing and future networks; 
\item[(2) {\sc Performance}:] the implementations of policies should be efficient on ``virtual switches'', that is with various resolutions from a single network element to the whole network (e.g., an interconnect for geographically distributed data centers~\cite{JainKMOPSVWZZZHSV13,HongKMZGNW13}); 
\item[(3) {\sc Simplicity}:] policies for different economical models should be expressible concisely, i.e., the code size and the organization of the code should be immediately reflective of the intent of the buffering infrastructure architect.
\end{inparaenum}

%
In this work we propose the \emph{Buffering Architecture SpEcification Language} (\barch{}), a flexible way to
meet these requirements and define buffering architectures and management policies that can be deployed on real network elements.


\section{Related Work}\label{sec:related-work}

The active networks~\cite{TennenhouseW96} approach to programmable networks is to execute code contained within packets on the switches. However, we argue that running arbitrary code can hamper switch performance.
Frenetic~\cite{FosterHFMRSW11}, Pyretic~\cite{MonsantoRFRW13}, and Maple~\cite{VoellmyWYFH13}, among others, have proposed 
abstractions to express management policies in packet networks. These approaches focus on abstractions for flexible \emph{classifiers}, and do not try to manage buffering architectures.
Other systems~\cite{SouleBMPKSF14,ShiehKGK10,FergusonGLFK13} allow for setting a \emph{predefined set} of parameters for
buffer management, which intrinsically limits expressivity. 
Another line of research abstracts the representation of the southbound API (e.g., OpenFlow) in the data plane~\cite{BosshartDGIMRSTVVW14,Song13,KozanitisHSV10}, while languages such as P4~\cite{BosshartDGIMRSTVVW14} are very successful in representing hierarchical tuple matching with action sets, we believe that they are less suitable to express buffer management policies. Usually queueing modules are physically separated from the \emph{packet processing engines}
(PPEs) implementing tuple-based classification~\cite{qfp,crs1}. This makes it difficult to access the current state of the queueing module. Conversely, adding classification to a queueing module can significantly increase the implementation cost, let alone add performance overheads.\footnote{A single TCAM access on Cisco C12000 linecard~\cite{pinnacle}
requires $11$ clock cycles, where the whole IP packet processing in one pipeline stage should be completed in $31$ cycles to guarantee the required line rate; running the second consecutive lookup that is based on dynamically changed values can degrade performance by up to $40\%$.}
We believe that buffer management policies should be built on different principles. 
The closest work to \barch{} is ~\cite{SivaramanWSB13} which introduces a set of primitives to define buffer management policies. While \barch{} only requires a set of comparators and conditions,~\cite{SivaramanWSB13} needs a specification
of the \emph{whole algorithm} for the management policy 
based its APIs, hence hampering simplicity. In addition, ~\cite{SivaramanWSB13} 
provides no clear separation from the classification module and the interface to express desired objectives. 
\barch{} aims to overcome these limitations.

\section{Buffering Architecture Design}\label{sec:overv-buff-arch}

The majority of buffering architectures can be defined with only two types of objects: \emph{ports}, and \emph{queues} 
assigned to ports; in the \emph{buffered crossbar} architecture~\cite{KeslassyKSS12}, cross-points
can also be represented as ports. An \emph{admission control policy}
for a queue determines which packets are admitted or
dropped~\cite{FloydJ93,FengSKS02,NicholsJ12}. A \emph{scheduling
  policy} for a port selects a queue whose \emph{head-of-line} (HOL)
packet will be processed next~\cite{DemersKS89,McKenney90}; in each
queue, the HOL packet (and thus the processing order) is defined by a \emph{processing policy}. 

In some cases, e.g., in a \emph{shared memory} switch~\cite{AielloKM08}, several queues share the same \emph{buffer} 
space, and admission control can routinely query the state of several queues; 
for instance, the Longest-Queue-Drop (LQD) policy drops packets from the longest queue in case of congestion~\cite{AielloKM08}. 
To cover these buffering architectures, we introduce one more object type, a \emph{buffer},
and an additional admission control policy to resolve congestions at the buffer level.
In a nutshell, to define a specific buffering architecture and its
management policy one creates instances of ports, queues, and
buffers, and specifies relations among them; admission control,
processing, and scheduling policies are attached to the corresponding
instances.
The purpose of \barch{} is to enable concise specification of buffering architectures and management policies.

\section{\barch{} Specification Language}\label{sec:barch-specification}

The abstractions introduced by \barch{} reconcile simplicity and expressivity.
\barch{} is a specification language, hence all programming decisions have to be made at the declaration of the different
entities, i.e., entities are static. 
In the following we shall present the different entities manipulated by \barch{} by means of a simple declaration of data structures (with no types). For each entity, we will define its properties, some of which are primitives of the domain (e.g., the size of a packet), and others which have to be set up by the programmer. Similarly, some properties are constant (e.g., maximum buffer capacity), whereas others are dynamic (e.g., the occupancy of a buffer). We show the characteristics of each property as comments. We denote by \lstinline|r| properties that are read-only, and by \lstinline|rw| properties that can also be updated by the user. Properties that are constant throughout the lifetime of the policy are marked with \lstinline|cons|, and with \lstinline|dyn| we mark properties that can change dynamically. Finally, functions are annotated with their return type (e.g., \lstinline|bool fun|). 



\subsection{Packets}
  
\begin{figure}[!t]
  \centering
  \begin{minipage}{0.8\linewidth}
\begin{lstlisting}[frame=tb,basicstyle=\ttfamily\small,belowskip=0em]
Packet {
 size       // size in bytes   [r, cons]
 value      // virtual value   [r, cons]
 processing // # of cycles      [r, dyn]
 arrival    // arrival time    [r, cons]
 slack      // offset in time  [r, cons]
 queue      // target queue id [r, cons]
}
\end{lstlisting}
    \caption{\barch{}'s packet primitive \label{list:buffer}}
  \end{minipage}
\end{figure}
\lstset{morekeywords={queue}}

In \barch{}, the notion of a packet is \emph{primitive}, meaning that the user cannot modify or extend packets; packet fields can be used to implement policies. Since a virtual switch can be defined with any resolution, from a real switch (or a part of it) to the entire network, and can represent the buffering architecture of different services, the notion of a packet is \emph{virtual} and is not related to specific traffic types. To be independent of traffic types and switch resolution, and to have a clear separation from the classification module, every incoming packet is prepended with three mandatory parameters -- an \emph{arrival} time, a packet \emph{size} in bytes, and a destination \emph{queue} -- and three optional parameters -- an intrinsic \emph{value} (whose
meaning is application-specific), the \emph{processing} requirement in virtual cycles, and \emph{slack} (maximal offset in time from \emph{arrival} when the packet must be transmitted). We assume that these properties  are set by an
external \emph{classification unit} (e.g., OpenFlow~\cite{OF}, if a virtual switch is defined with the finest possible resolution), except for \emph{arrival} (which is set by \barch{} when a packet is received) and \emph{size}.

Figure~\ref{list:buffer} shows an abstract representation of the \lstinline|Packet| data structure.
Intrinsic \emph{value} and \emph{processing} requirements can be useful to define prioritization
levels~\cite{KeslassyKSS12}. The \emph{slack} represents a time bound, which can be used in management decisions
of latency-sensitive applications; for instance, if buffer occupancy already exceeds the \emph{slack} value of an
incoming packet, the packet can be dropped during admission even if there is available buffer space. In Section~\ref{sec:examples}, we will see specific examples that exploit these characteristics.

We postulate that all decisions of buffer management policies (during admission or scheduling) are based only on the
specified packet parameters and internal state variables of a buffering architecture (e.g., buffer occupancy).

\begin{figure}[t!]
\centering
  \begin{minipage}{1.04\linewidth}
\begin{lstlisting}[frame=tb,basicstyle=\ttfamily\small,belowskip=0em]
Queue {
 // primitive properties
 currSize       // current size             [r, dyn]
 getHOL()       // head-of-line pkt     [packet fun]

 // user-specified at declaration
 size           // size in bytes           [r, cons]
 buffer         // buffer where allocated  [r, cons] 
 procPrio(p1,p2)// processing prio comp.  [bool fun]
 admPrio(p1,p2) // pushOut prio comp.     [bool fun]
 congestion()   // congestion predicate   [bool fun]
 postAdmAct()   // {MARK,NOTIFY,..}     [action fun]
 weightAdm      // priority for admission  [rw, dyn]
 weightSched    // priority for scheduling [rw, dyn]
}
\end{lstlisting}
    \caption{\barch{}'s queue primitive\label{list:queue}}
  \end{minipage}
  \end{figure}

\subsection{Queues}

Figure~\ref{list:queue} summarizes the API provided to the programmer to declare queues.  The standard property \lstinline|size| is defined by the user at declaration time.
The \lstinline|currSize| property changes dynamically as the queue changes its size.  
Abstractly, a queue contains packets ordered according to user-defined priorities for admission control 
and processing policies. In \barch{}, we consider two user-defined priorities:

(a)
\lstinline|procPrio(p1,p2)| is a Boolean function that takes
  two abstract packets and returns \emph{true} only if
  \lstinline|p1| has a higher processing priority than \lstinline|p2|. We call functions that compare any two
  objects of the same type \emph{comparators} (defined as Boolean expressions with arithmetic/Boolean operators
  and access to packet and object attributes),
so \lstinline|procPrio| is a packet comparator. In \barch{} we are only concerned with the highest processing priority packet at any point. Hence, the only way to access the queue ordered by \lstinline|procPrio| is through the \lstinline|getHOL()| primitive which returns the HOL (i.e., highest processing priority as defined by \lstinline|procPrio|) packet in the queue. E.g., 
to encode a FIFO processing priority the user sets
\lstinline[basicstyle=\ttfamily\small]|procPrio(p1,p2) = p1.arrival < p2.arrival|. \newline
With this definition, each call to \lstinline|getHOL()| returns the packet in the queue with the oldest arrival time.

(b)
\lstinline|admPrio(p1,p2)| is also a packet comparator used in case of congestion to
choose the packets that should be dropped from the queue to restore the queue into a decongested state.
We could have simply chosen to use the least valuable packets according to \lstinline|procPrio| for drops,
but we will see in Sec.~\ref{sec:examples} that separate priorities for admission and processing 
has more flexibility and improves performance. 

To indicate when a queue is virtually \emph{congested}, we use a user-defined predicate \lstinline|congestion()|. 
The optional function \lstinline|postAdmAct()| returns an action applied after admission and
can update \lstinline|weightAdm| (if necessary). The function \lstinline|postAdmAct()| can also be used to implement \emph{explicit congestion notifications}~\cite{BauerBB11} or \emph{backpressure}; \lstinline|postAdmAct()| can return actions as \lstinline|MARK|, \lstinline|NOTIFY|, etc. For cases when bandwidth is allocated not only with 
respect to packet attributes, queues maintain a \lstinline|weightSched| variable that can be updated dynamically
after each scheduling operation. With \lstinline|weightSched| one can, e.g., define static
bandwidth allocation among queues of the same port during scheduling decisions; \lstinline|weightSched| can be 
updated in the \lstinline|postSchedAct()|  function which is defined at the port level.



\subsection{Ports}

\begin{figure}[t]
	\centering
  \begin{minipage}{.938\linewidth}
\begin{lstlisting}[frame=tb,basicstyle=\ttfamily\small,belowskip=0em]
Port {
 // primitive properties
 getBestQueue() // on weightSched   [queue fun]
 getCurrQueue() // scheduled one    [queue fun]

 // user-specified at declaration
 schedPrio(q1,q2)// compare q-s      [bool fun]
 postSchedAct() //{MARK,NOTIFY,..} [action fun]
}
\end{lstlisting}
      \caption{\barch{}'s port primitive}\label{list:ports}
  \end{minipage}
\end{figure}

The interface provided for ports is presented in Figure~\ref{list:ports}.
Each port manages a set of queues assigned to the port at its declaration.\footnote{We leave the \lstinline{new}
operator used to create network objects in \barch{} implicit; its usage will be clear from the examples in Sec.~\ref{sec:examples}.}
The policy \lstinline|schedPrio(q1,q2)| is a user-defined (queue comparator) scheduling property that defines which HOL packet is scheduled next (this packet is accessed through the \lstinline|getBestQueue()| function). 
For example, a priority based on packet values which implements several levels of strict priorities is declared as follows:

\begin{lstlisting}[basicstyle=\tt\footnotesize]
schedPrio(q1,q2) = 
       q1.getHOL().value > q2.getHOL().value
\end{lstlisting}

Finally, \lstinline|postSchedAct()| is similar to the \lstinline|postAdmAct()| function of queues which can be used to define new services.



\begin{figure}[t!]
  \centering
 \begin{minipage}{.9\linewidth}
\begin{lstlisting}[frame=tb,basicstyle=\ttfamily\small,belowskip=0em]
Buffer {
 // primitive properties
 currSize       // current size      [r, dyn]
 getBestQueue() // on weightAdm   [queue fun]
 getCurrQueue() // admitted one   [queue fun]
   
 // user-specified at declaration
 size           // size             [r, cons]
 congestion()   // cong. predic.   [bool fun]
 queuePrio(q1,q2)// compare q-s    [bool fun]
 postAdmAct() //{MARK,NOTIFY,..} [action fun]
}
\end{lstlisting}
      \caption{\barch{}'s buffer primitive}\label{list:buffers}
  \end{minipage}
\end{figure}

\subsection{Buffers}
 The interface provided for declaring buffers is presented in Figure~\ref{list:buffers}.
A buffer is an optional entity; it is declared only in the case when several queues share buffer space. Each buffer manages a set of queues assigned to it at creation time; \lstinline|congestion()|, \lstinline|postAdmAct()|, \lstinline|size|, and \lstinline|currSize| are similar to the corresponding queue attributes. In case of congestion, an admission control policy on the buffer level finds a queue whose packet should be dropped, and the admission control policy of the chosen queue determines which packet is dropped. To order queues for admission, the user specifies the \lstinline|queuePrio| comparator. For instance, to implement LQD 
the following comparator can be used:

\begin{lstlisting}[basicstyle=\ttfamily\footnotesize]
queuePrio(q1,q2) = q1.currSize < q2.currSize
\end{lstlisting}

\section{\barch{} at Work (examples)}\label{sec:examples}


\begin{figure}
\centerline{
\includegraphics[scale=.3]{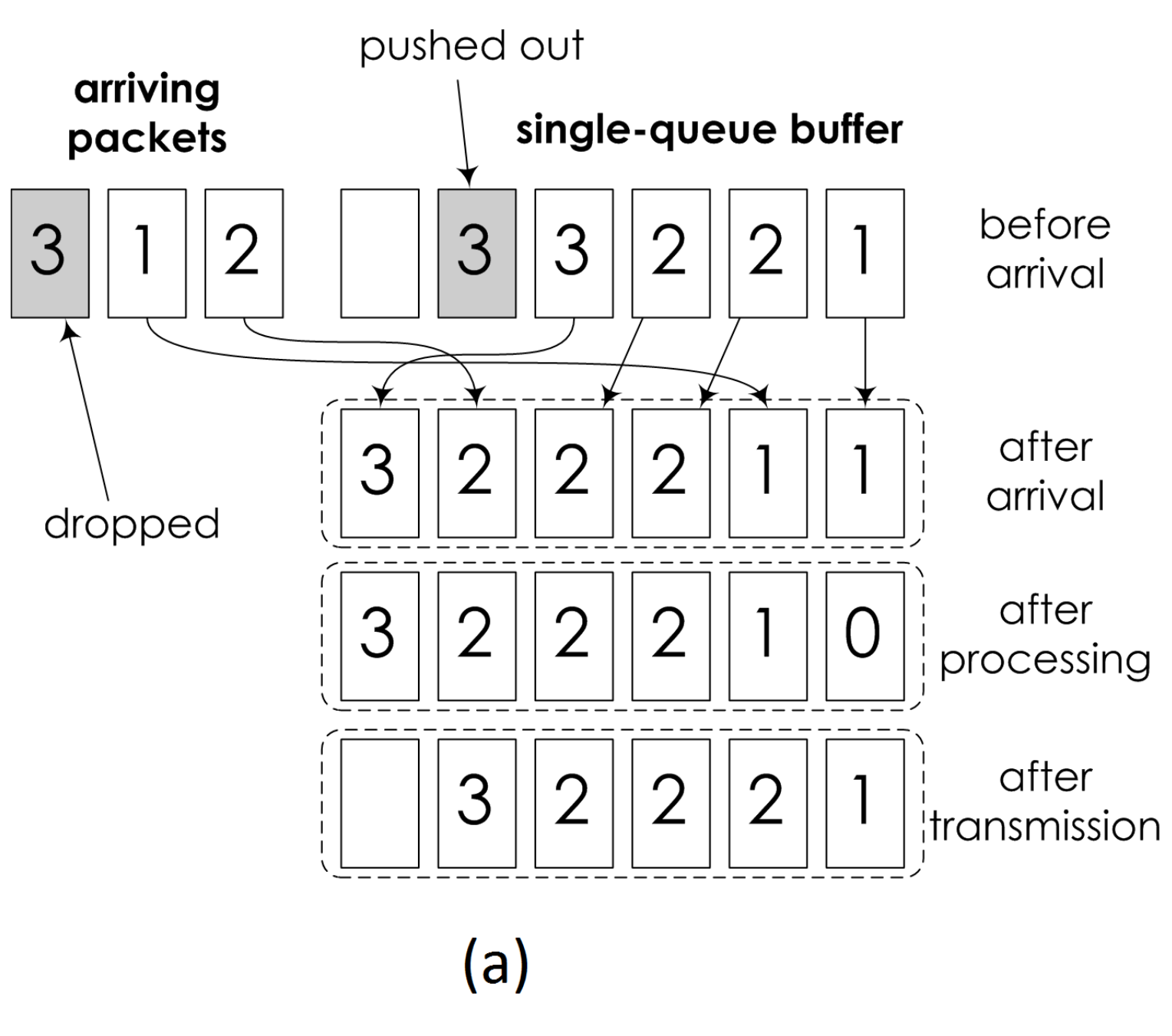}
\includegraphics[scale=.3]{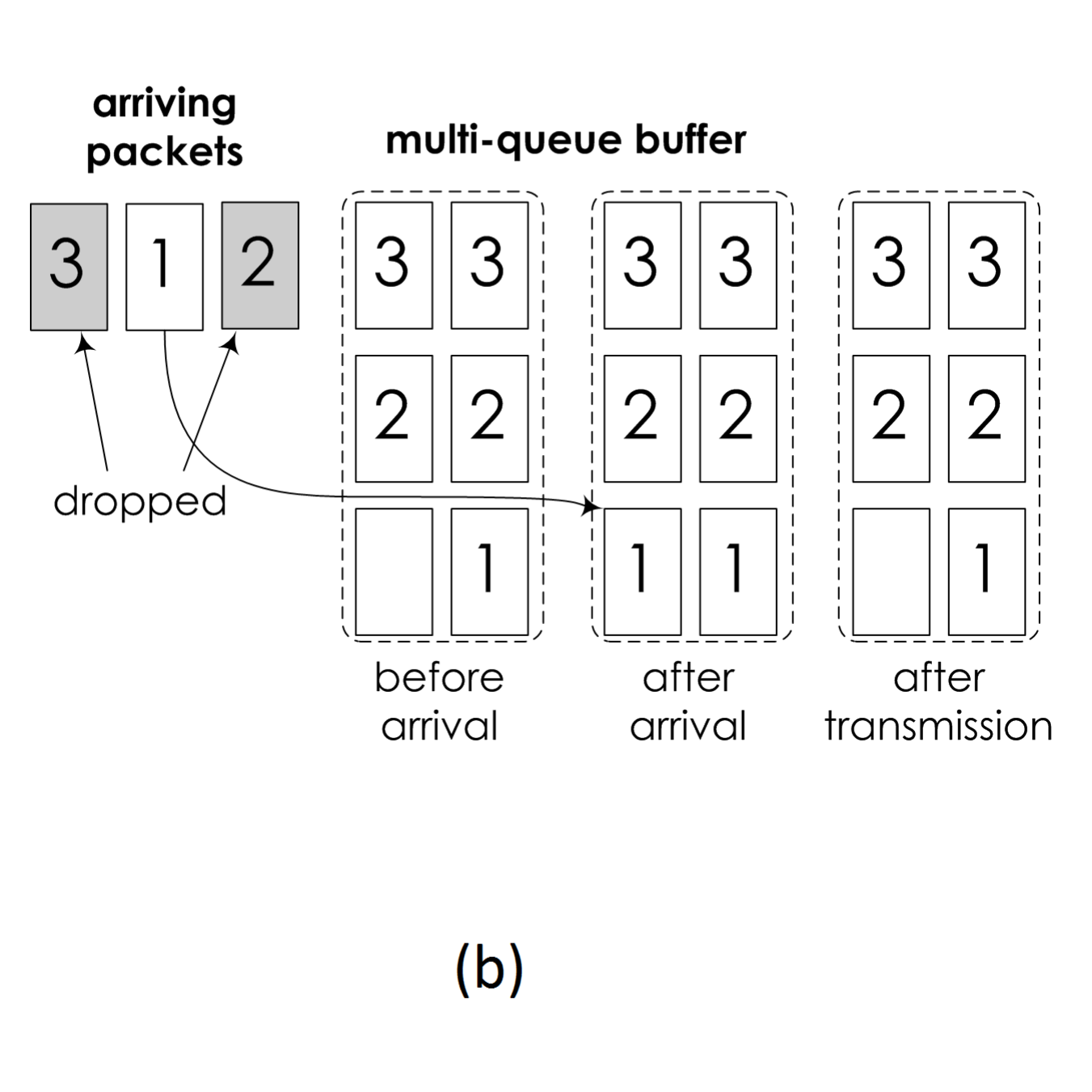}
}
\vspace{-10pt}
\caption{Left: single priority queue with buffer $B = 6$; right: multi-queued switch with three queues ($k = 3$) and buffer $B = 2$ each. Dashed lines enclose queues.}
\label{fig:model}
\end{figure}

\newcommand{\ttfam}[1]{{\ttfamily\footnotesize #1}}

\begin{table} \centering
\small
\vspace{2mm}
	\lstset{basicstyle=\ttfamily\footnotesize}
\begin{tabular}{ccc}\hline
\ttfam{admPrio}  & \ttfam{procPrio} & OPT/ALG \\\hline
\ttfam{fifo()} & \ttfam{fifo()}  & $O(k)$ \\
\ttfam{fifo()} & \ttfam{srpt()}  & $O(\log{k})$ \\
\ttfam{rsrpt()} & \ttfam{srpt()} & $1$ (optimal) \\\hline
\end{tabular}
\vspace{2pt}
\caption{
  Sample \barch{} policies with analytic results, single queue architecture;
  $k$ is the maximal processing requirement, OPT/ALG
  is the competitive ratio.
  }\label{tbl:single-queue}
\end{table}


To demonstrate the impact of admission and processing orders on the efficiency of admission control, consider throughput maximization in a single queue buffering architecture (buffer of size $B$), each unit-sized and unit-valued packet assigned with a number of required processing cycles ranging from $1$ to $k$ (see Figure~\ref{fig:model}(a)).
Figure~\ref{list:inst} shows a sample definition in \barch{} for a single queue buffering architecture, where we
deploy a single \lstinline|q1| in a single port \lstinline|out|. Below we will see how to define different
management policies in this architecture. In this example, the \lstinline|fifo()| packet comparator uses
arrival time, the \lstinline|srpt()| (shortest remaining processing time) and \lstinline|rsrpt()| (reversed shortest remaining processing time) comparators use remaining processing requirements. The congestion condition
declared in \lstinline|defCongestion()| is trivial, satisfied when
occupancy exceeds queue size. We instantiate a single queue \lstinline|q1| 
with this congestion policy. 

\newcommand{\ttfamsm}[1]{{\ttfamily\small #1}}

\begin{figure}[h]
\begin{lstlisting}[frame=tb,
	basicstyle=\ttfamily\small]
// considered priorities for admission and 
// processing
fifo(p1,p2) = (p1.arrival < p2.arrival)
srpt(p1,p2) = (p1.processing < p2.processing)
rsrpt(p1,p2) = (p1.processing > p2.processing)

// default congestion condition for all 
// considered policies
defCongestion() = lambda q, (q.currSize >= q.size)

// initializing a generic buffering architecture
q1=Queue(B); out=Port(q1);
q1.proPrio(p1,p2)=fifo(p1,p2);
q1.congestion=defCongestion(q1);
\end{lstlisting}
\vspace{-5pt}
\caption{Single queue buffering architecture in \barch{}}\label{list:inst}
\end{figure}

Table~\ref{tbl:single-queue} lists implementations for \lstinline|admPrio| and \lstinline|procPrio| in this architecture and analytic competitiveness results for various online policies versus the optimal offline algorithm~\cite{KeslassyKSS12,NikolenkoK15}.
Each row represents a management algorithm for a single queue; e.g., the first row shows a simple greedy algorithm that admits every incoming packet if possible (see \lstinline|congestion()|), and processes them in \lstinline|fifo()| order; it is $O(k)$-competitive for maximum processing requirement $k$. In \barch{}, this algorithm looks as follows:

\begin{lstlisting}[basicstyle=\ttfamily\footnotesize]
q1.admPrio=fifo; q1.procPrio=fifo;
\end{lstlisting}

\noindent
On the other hand, changing the processing order \lstinline|fifo()| to
\lstinline|srpt()| introduces a significant improvement in performance
and this version of the greedy policy is already $O(\log{(k)})$-competitive. 
With the third greedy algorithm that processes packets in \lstinline|srpt()| order and admits them in \lstinline|rsrpt()| order, we get an optimal algorithm for throughput maximization regardless of traffic distributions~\cite{KeslassyKSS12}. Since here a port manages only one queue, a \emph{scheduling policy} is just an implicit call to \lstinline|getHOL()|.

One alternative architecture for packets with heterogeneous processing
requirements is to allocate queues for packets with the same processing requirements (see Figure~\ref{fig:model}(b)). 
The following code creates this buffering architecture in \barch{},
where \lstinline|k| queues share an equal portion of memory \lstinline|B/k|.

\begin{lstlisting}[basicstyle=\ttfamily\footnotesize]
// creation of buffering architecture
q1=Queue(B/k);...qk=Queue(B/k); 
out=Port(q1,..,qk);
\end{lstlisting}

In this architecture, there is no need for advanced processing and admission orders since only packets with the same processing requirement are admitted in the same queue. 
The following \barch{} code instantiates \lstinline|admPrio|,\linebreak \lstinline|procPrio| and \lstinline|congestion|  in the $k$ created  queues.

\begin{lstlisting}[basicstyle=\ttfamily\footnotesize]
q1.admPrio=fifo; ...; qk.admPrio=fifo;
q1.procPrio=fifo; ...; qk.procPrio=fifo;
q1.congestion=defCongestion(q1); ...; 
qk.congestion=defCongestion(qk);
\end{lstlisting}

This change of buffering architecture is not for free since the buffer of these queues is not shareable. But even here, the decision of which packet should be processed to maximize throughput is non-trivial since it is unclear which characteristic is most relevant for throughput optimization: buffer occupancy, required processing, or a combination. 
\barch{} code in Figure~\ref{list:priosandactions} presents six different scheduling priorities and \lstinline|postSchedAct| actions in the cases when this action is used. 

\begin{figure}[h]
\begin{lstlisting}[frame=tb,basicstyle=\ttfamily\small,deletekeywords={packet,queue},belowskip=0em]
// LQF: HOL packet from Longest-Queue-First
lqf(q1,q2)  = (q1.currSize > q2.currSize);
// SQF: HOL packet from Shortest-Queue-First
sqf(q1,q2)  = (q1.currSize < q2.currSize);
// MAXQF: HOL packet from queue that 
// admits max processing
maxqf(q1,q2)= (q1.weightSched > q2.weightSched);
// MINQF: HOL packet from queue that admits 
// min processing
minqf(q1,q2)= (q1.weightSched < q2.weightSched);
// CRR: Round-Robin with per cycle resolution
crr(q1,q2)  = (q1.weightSched < q2.weightSched);
crrPostSchedAct() = lambda port, 
         (port.getCurrQueue().weightSched += k);
// PRR: Round-Robin with per packet resolution
prr(q1,q2)  = (q1.weightSched < q2.weightSched);
prrPostSchedAct() = lambda port, 
  (let q = port.getCurrQueue() in 
    if (q.getHOL().processing == 0) 
        q.weightSched += k*k));
\end{lstlisting}
\caption{\barch{} example of scheduling priorities and \lstinline|postSchedAct| actions for multiple separated queues.}\label{list:priosandactions}
\end{figure}

\begin{figure}
\centerline{
\includegraphics[scale=0.35]{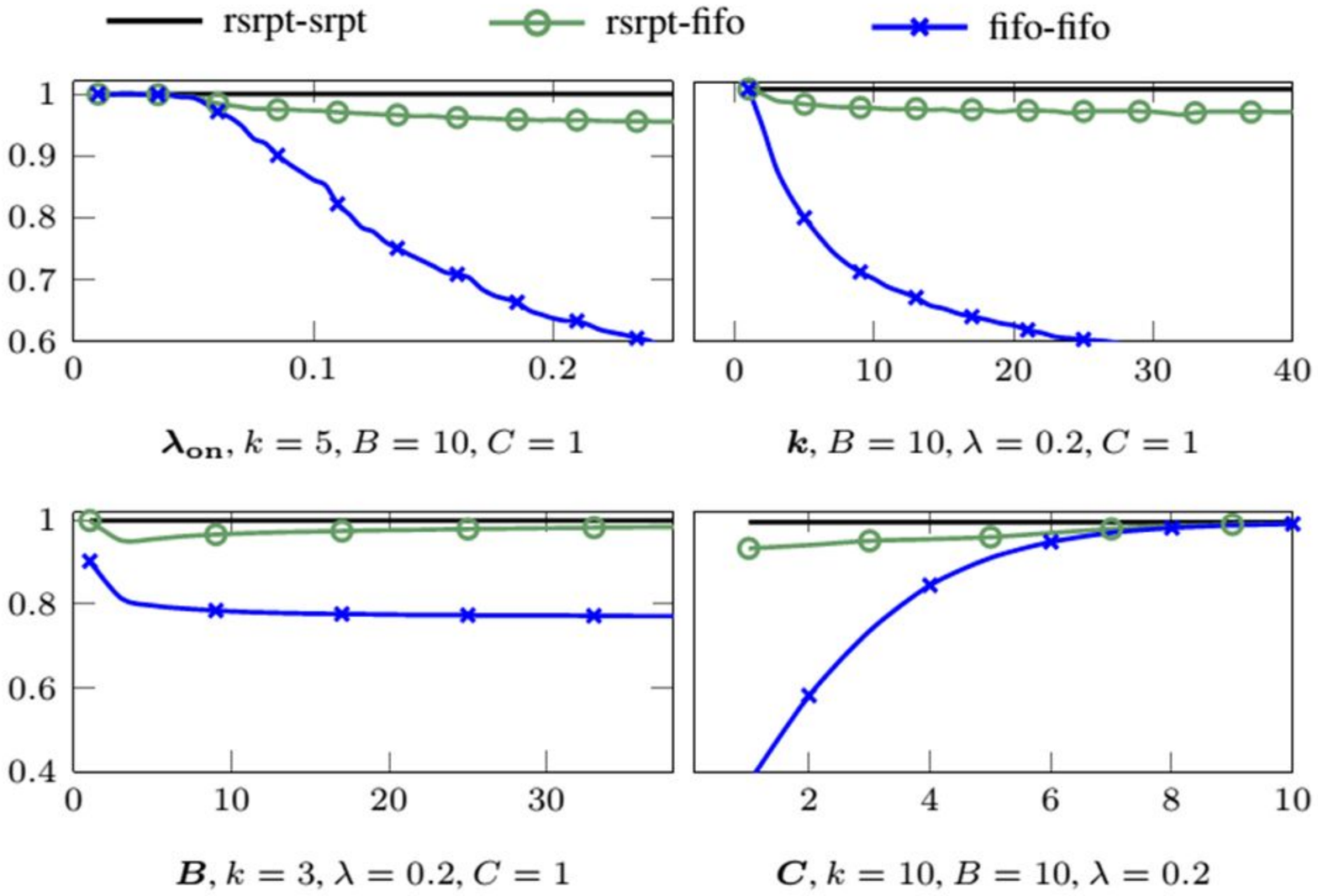}
}
\vspace{-10pt}
\caption{Optimal vs three online algorithms for a single queue architecture with heterogeneous processing;
$y$-axis, competitive ratio; $x$-axis, top to bottom, left to right: $\lambda$; max required processing $k$; 
buffer size $B$; speedup $C$.}
\label{fig:sim-queue}
\end{figure}

\begin{figure}
\centerline{
\includegraphics[scale=0.35]{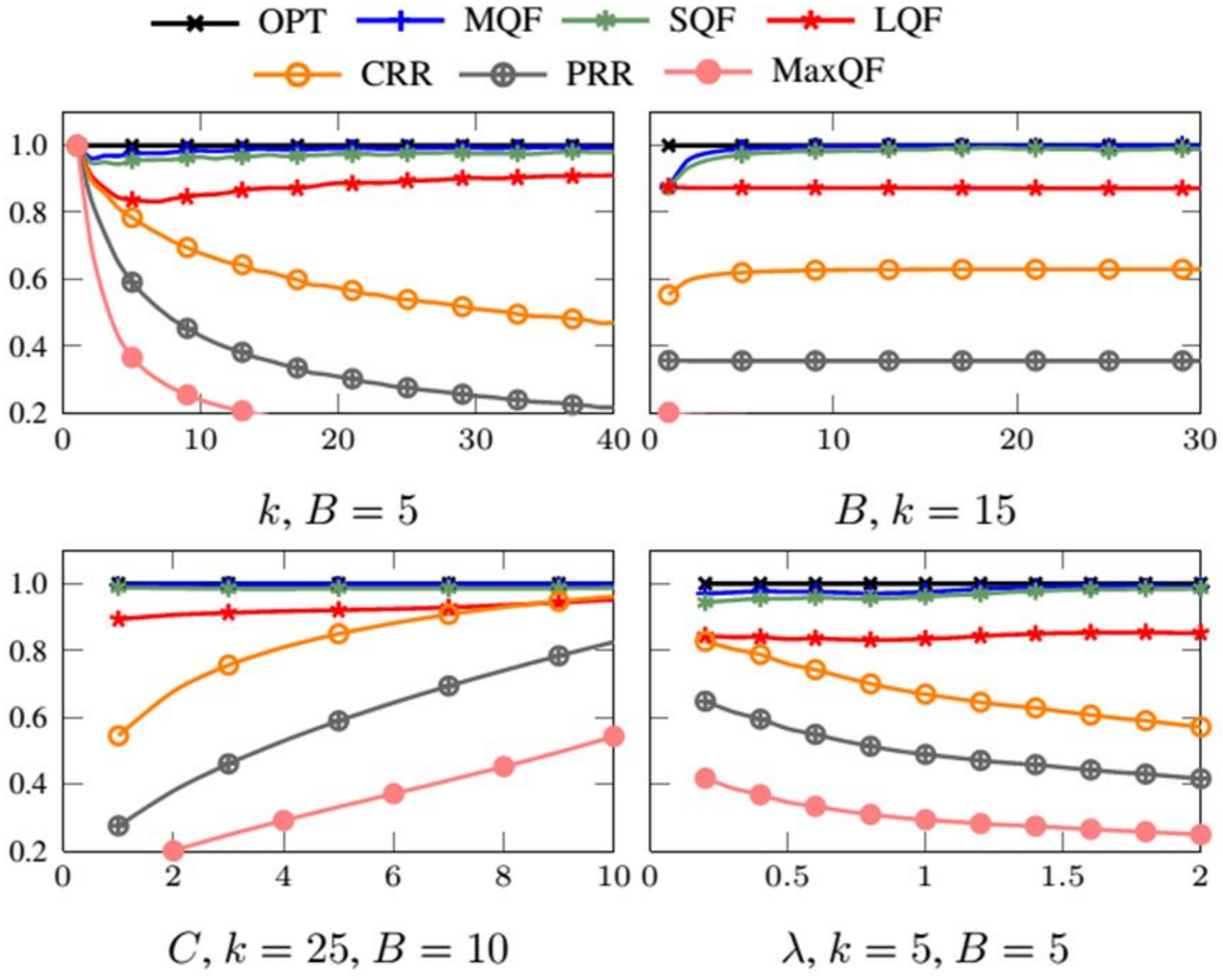}
}
\vspace{-10pt}
\caption{Online vs optimal algorithms for multiple que\-ues with heterogeneous processing;
$y$-axis, competitive ratios; $x$-axis, top to bottom, left to right: max required processing $k$, buffer size $B$, speedup $C$, intensity $\lambda$.}
\label{fig:sim-queues}
\end{figure}

\begin{table} \centering
\small
\lstset{basicstyle=\ttfamily\small}
\setlength{\tabcolsep}{2pt}\renewcommand{\arraystretch}{1.5}
\begin{tabular}{cccc}\hline
init. \ttfamsm{weightSched} & \ttfamsm{postSchedAct} & \kern-8pt\ttfamsm{schedPrio}  & OPT/ALG \\\hline
unused & unused & \ttfamsm{lqf()} & $\Omega({\frac{B}{2}})$ \\
unused & unused & \ttfamsm{sqf()} & $\Omega(k)$\\
unused & unused & \ttfamsm{maxqf()} & $\Omega(k)$\\
\kern-5pt
\ttfamsm{qi.w}eightSched=i & unused & \ttfamsm{minqf()} &
                                                              \kern-3pt upper bound $2$ \\
\kern-5pt
\ttfamsm{qi.w}eightSched=i & \ttfamsm{crrPostSchedAct()} & \ttfamsm{crr()}	& $\Omega({\frac{k}{\ln{k}}})$ \\[2pt]
\kern-5pt
\renewcommand{\arraystretch}{1}
\ttfamsm{qi.w}eightSched=i & \ttfamsm{prrPostSchedAct()} & \ttfamsm{prr()} & $\Omega({\frac{3k(k+2)}{4k+16}})$\\\hline
\end{tabular}
\vspace{-6pt}
\caption{Examples of policies in \barch{} for multiple queues architecture with analytic results; $k$ is a maximal processing requirements, $B$ is a buffer size of a single queue. OPT/ALG is a throughtput of an optimal offline OPT algorithm vs. online algorithm ALG.}\label{tbl:multiple}
\end{table}
Table~\ref{tbl:multiple} summarizes various online scheduling policies as shown in~\cite{KoganLNS13,NikolenkoK15}.
Observe that buffer occupancy is not a good characteristic for throughput maximization: \ttfamsm{lqf()} and 
\ttfamsm{sqf()} have bad competitive ratios, while a simple greedy scheduling policy Min-Queue-First (MQF) that
processes the HOL packet from the non-empty queue with minimal required processing (\ttfamsm{minqf()}) is 
2-competitive. This means that MQF will have optimal throughput with a moderate speedup of $2$~\cite{KoganLNS13}. 
The other two policies that implement fairness with per-cycle or per-packet resolution (CRR and PRR respectively) 
have relatively weak performance; this demonstrates the fundamental tradeoff between fairness and throughput.
The following code snippet in \barch{}, for instance, corresponds to the CRR policy:

\begin{lstlisting}[basicstyle=\ttfamily\footnotesize]
// initializing schedWeight for CRR
q1.weightSched=1; ... qk.weightSched=k;
// initial. postSchedAct to update schedWeight
out.postSchedAct = crrPostSchedAct(out); 
\end{lstlisting}

Currently, the best tools available to evaluate performance of buffering architectures are discrete simulators such as 
NS-2~\cite{ns2} or OMNet++~\cite{omnnet} that can use traffic traces and/or various traffic distributions to 
analyze performance of buffering architectures by specifying management policies in a high level language. 
Due to its simplicity, \barch{} can be used as a discrete simulator whose configuration is limited to several
user-defined expressions. 
For instance,
Figures~\ref{fig:sim-queues} and~\ref{fig:sim-queue} show the impact
of admission, processing, and scheduling policies on throughput
optimization for a single queue and multiple queues buffering
architectures with packets of heterogeneous processing requirements;
in these examples, traffic was generated with an ON-OFF
Markov modulated Poisson process (MMPP) with Poisson arrival processes
with intensity $\lambda$, and required processing chosen uniformly at
random from $1..k$. But even if we know how to represent arrivals and
analyze them, the applicability of these results will be limited to
specific settings. Hence, \barch{} is being developed for deployment on real
systems.

\def\legendsize{\scriptsize}

\section{\barch{} Internals}\label{sec:implementation}
One of the fundamental building blocks in \barch{} is the \emph{priority queue} data structure where the order of elements is based on a user-defined priority. The implementation keeps a single copy of the packets, and priority queues are implemented with pointers to actual packets. Since a virtual buffering architecture can be defined with switch resolution (or a part of it), \barch{} implementation is reduced to efficient implementation of a priority queue data structure that can operate at line rate. While in the general case priority queue operations take $O(\log(N))$ time, where $N$ is its size, there are restricted versions (e.g., for a predefined range of priorities) that can support most operations in $O(1)$ and can be efficiently implemented even in hardware~\cite{MortonLS07,IoannouK07}.
To guarantee a constant number of insert/remove and lookup operations during admission or scheduling of a packet (i.e., to avoid rebuilding the priority queue), user-defined expressions for priorities are fixed during operation.


A \emph{push-out mechanism} makes an architecture capable to push out already admitted packets. This mechanism is supported in \barch{}. To avoid different implementations for the push-out and non-push-out cases, an admission control policy always virtually admits an incoming packet. In the event of a virtual congestion, admission control drops the least valuable packets until congestion is lifted.
The complexity of \barch{} is reduced to translating user-defined settings to a target system that implements a virtual buffering architecture. In some cases, the target should be extended, or the expressiveness of \barch{} can be restricted. 
We are currently implementing \barch{} on top of the Open vSwitch (OVS) as a sample target
architecture with the finest resolution~\cite{ovs,PfaffPKJZRGWSSA15}.

%

\subsection{\barch{} implementation in Open vSwitch}

\begin{figure}[t]
\centering
\hspace{-2mm}
\includegraphics[scale=.5]{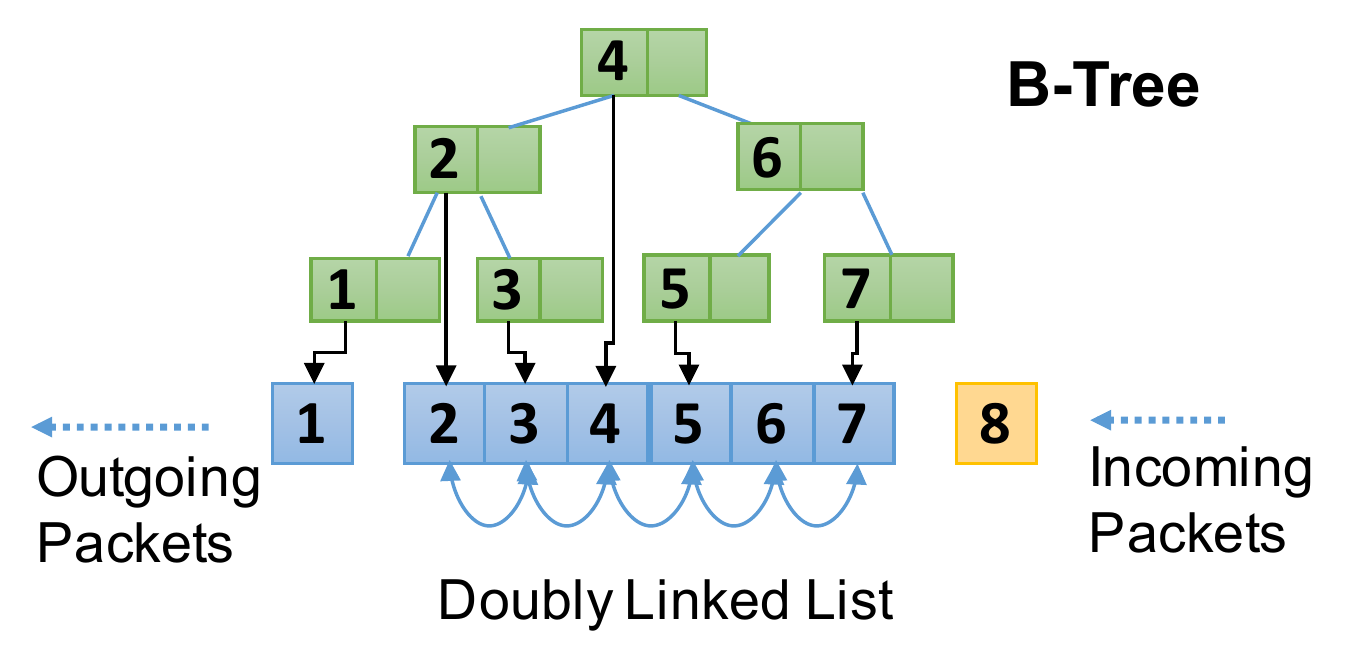}
\vspace{-2.5\topsep}
\caption{Priority queue implementation.}
\vspace{-6pt}
\label{fig:priority_queue}
\end{figure}

\begin{figure}[t]
\centering
\hspace{-2mm}
\includegraphics[scale=.42]{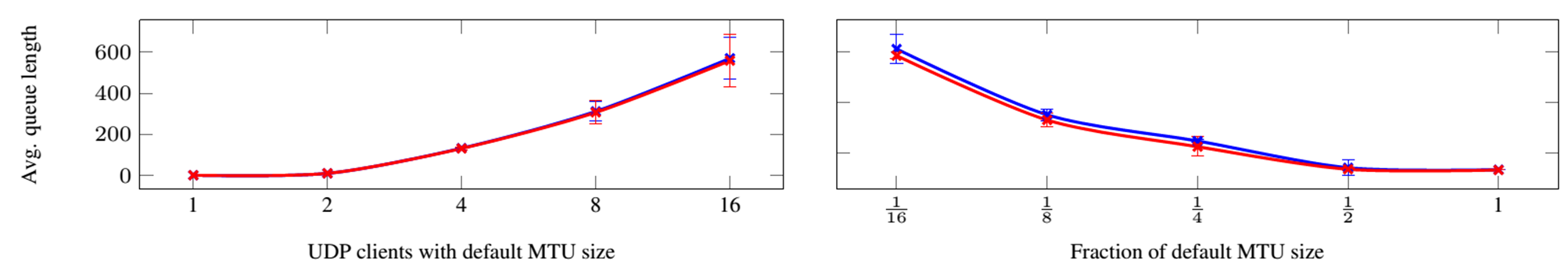}
\caption{\emph{Left:} average queue length as a function of number of clients generating UDP traffic with default MTU size. \emph{Right:} fraction of default MTU size; blue: FIFO with prioritization; red: regular FIFO.}
\vspace{-6pt}
\label{fig:avg_queue_length}
\end{figure}

OVS implements the control plane in user space and the data plane in the kernel.
To support \barch{} on the control plane, we extend OVSDB (Open vSwitch Database) with the notions of port, queue, and buffer described in Section~\ref{sec:barch-specification}. Moreover, since OVS exploits Linux TC (Traffic Control) kernel modules via the \verb|netdev-linux| library to manipulate queuing and scheduling disciplines (\verb|qdisc|), we are also adding configuration options to TC to express \barch{}'s admission, processing, and scheduling policies.
Similar extensions are being added on the data plane via Linux kernel TC loadable kernel modules.
\subsection{\barch{} feasibility}
We have extended Linux's default qdisc\footnote{Queuing Discipline (qdisc) is an integral part of Linux Traffic Controlling (TC) used to shape outgoing (egress) traffic for an interface; qdisc has an enqueue method to handle outgoing packets and a dequeue method to fetch packets written to the network interface.} (i.e., pfifo\_fast) to support packet prioritization based on arrival time. Instead of modifying the underlying default packet queue (a doubly linked list), we use an existing B-Tree implementation on top of a default FIFO queue to manage packet prioritization to preserve backward compatibility to existing qdisc solutions. As shown on Figure~\ref{fig:priority_queue}, we add a reference to the enqueuing packets to the B-Tree and the highest priority packet (i.e., the earlest arrival time) is dequeued first. We remark that FIFO does not need to utilize a B-Tree in general; we use it as a baseline to explore the performance overhead of a generic implementation of prioritization. 

In our testbed, we set a 3-node line topology to measure the performance overhead of our packet prioritization logic. Figure~\ref{fig:testbed} shows that the middle node runs OVS with modified data plane
(Linux kernel) and acts as a pass-through switch. We vary the number of parallel traffic generators on the first node
and measure average queue length (i.e., number of packets in the default queue) in a receiver node on the third
for two qdiscs: default FIFO and extended FIFO with prioritization, reporting the average value of $50$ runs with
$95$\% confidence interval. Figure~\ref{fig:avg_queue_length}(left) shows the average queue lengths for the
two qdiscs; in both cases, average queue length increases with the number of UDP clients. In FIFO with $16$ clients,
the most congested case, regular FIFO has average queue length $559.333$ vs. $571$ for FIFO with prioritization,
only a $2\%$ degradation. 
We also varied MTU sizes in the same 3-node line topology testbed with $4$ parallel UDP generators, which is a good enough case to observe queue build-ups but not dropping packets in the pass-though switch. We measured average queue lengths of the two qdiscs by varying MTU sizes from $\frac1{16}$ of the default MTU size to its default size ($1500$ bytes). Figure~\ref{fig:avg_queue_length}(right) shows that for both qdiscs the average queue length decreases as MTU size increases; FIFO with prioritization incurs only $4$\% overhead: for MTU size of $\frac{1500}{16}$ bytes the result is $584.3$ vs. $610.7$. Hence, we conclude that packet prioritization on top of FIFO incurs negligible performance overhead.



\begin{figure} 
\end{figure}

\vspace{-5pt}
\section{Conclusion}\label{sec:conclusion}
We propose a simple yet expressive language to define buffering architectures and their management policies. The proposed language is independent from a classification module and can define buffering architectures and their management policies with any resolution from a single network element to a virtual switch that can represent the whole network or a part of it. We believe that the efficient representation of buffering architectures in \barch{} can enable and accelerate innovation in this domain.

%

\label{last-page}

\end{document}